\documentclass[aps,preprint,showpacs,groupedaddress,floatfix]{revtex4}

\usepackage{graphicx}
\usepackage{dcolumn}
\usepackage{bm}

\newcommand{\bsr}[1][r]{\boldsymbol{#1}}

\begin{document}

\bibliographystyle{prsty}
\title{Relativistic QRPA calculation of total muon capture rates}
\author{T. Marketin}
\author{N. Paar}
\author{T. Nik\v si\' c}
\author{D. Vretenar}
\affiliation{Physics Department, Faculty of Science, University of Zagreb, 
Croatia}
\date{\today}
\begin{abstract}
The relativistic proton-neutron quasiparticle random phase approximation 
(PN-RQRPA) is applied in the calculation of total muon capture rates on a 
large set of nuclei from $^{12}$C to $^{244}$Pu, for which experimental values 
are available. The microscopic theoretical framework is based on the Relativistic 
Hartree-Bogoliubov (RHB) model for the nuclear ground state, and transitions 
to excited states are calculated using the PN-RQRPA. The calculation is fully 
consistent, i.e., the same interactions are used both in the RHB equations that 
determine the quasiparticle basis, and in the matrix equations of the PN-RQRPA.
The calculated capture rates are sensitive to the in-medium quenching 
of the axial-vector coupling constant. By reducing this constant from its 
free-nucleon value $g_A = 1.262$ by 10\% for all multipole transitions, the 
calculation reproduces  the experimental muon capture rates to better than 
10\% accuracy.
\end{abstract}
\pacs{ 21.60.Jz, 23.40.Bw, 24.30.Cz, 25.30.Mr}
\maketitle
\date{today}
\section{Introduction}

Semi-leptonic weak interaction processes in nuclei are very sensitive to 
detailed properties of nuclear ground states and excitations. In astrophysical 
applications, in particular, weak interaction rates ($\beta$-decay half lives, 
neutrino-nucleus cross sections, electron capture rates) must be calculated for 
hundreds of isotopes. Many of those are located far from the valley of $\beta$-stability, 
and thus not easily accessible in experiments. For a consistent description, 
reliable predictions and extrapolations of these processes it is, therefore, essential 
to employ a consistent theoretical framework based on microscopic nuclear 
structure models.

At present the framework of nuclear energy density functionals (NEDF) provides 
the most complete description of ground-state properties and 
collective excitations over the whole nuclide chart. At the level of practical 
applications the NEDF framework is realized in terms of self-consistent 
mean-field (SCMF) models. With a small set of universal parameters 
adjusted to data, the SCMF approach has achieved a high level 
of accuracy in the description of structure properties over the whole 
chart of nuclides, from relatively light systems to superheavy nuclei, and from 
the valley of $\beta$-stability to the particle drip-lines \cite{BHR.03,Vretenar2005a}.

In a series of recent studies we have used a fully consistent microscopic 
approach based on relativistic energy density 
functionals to analyze $\beta$-decay half-lives of neutron-rich nuclei \cite{Niksic2005,Marketin2006}, and to model
inclusive charged-current neutrino-nucleus reactions \cite{Paar2008}. 
In this framework nuclear 
ground states are described using the relativistic Hartree-Bogoliubov (RHB) model 
\cite{Vretenar2005a}, and transitions to excited nuclear states are calculated 
in the relativistic quasiparticle random-phase approximation (RQRPA)
\cite{Paar2003,Paar2004}. 
There are important advantages in using functionals with manifest 
covariance, the most obvious being the natural inclusion of the 
nucleon spin degree of freedom. The resulting nuclear spin-orbit 
potential has the correct empirical strength and isospin dependence. 
This is, of course, 
especially important in the description of excitations in the spin-isospin 
channel, e.g. semi-leptonic weak interaction processes. In addition, by employing  
a single universal effective interaction in modeling both 
ground-state properties and multipole excitations in various mass 
regions of the chart of nuclides, the calculation of weak-interaction 
rates is essentially parameter free, and can be extended 
to regions of nuclei far from stability, including those on 
the $r$-process path.

To successfully extend a particular microscopic approach to regions of unknown 
nuclei far from stability, it is necessary to perform extensive tests and compare 
results with available data. Reliable prediction of weak
interaction rates, in particular, require a fully consistent description 
of the structure of ground states and multipole excitations. For instance, 
calculated $\beta$-decay half-lives are very sensitive to low-energy 
Gamow-Teller transitions, but can only test excitations of 
lowest multipoles. Higher multipoles are excited in neutrino-nucleus
reactions in the low-energy range below 100 MeV, and these reactions
could play an important role in many astrophysical processes, including
stellar nucleosynthesis. There are, however, only few data on neutrino-nucleus 
reactions, and these are limited to relatively light nuclei. Much more data 
are available for total muon capture rates. Muon capture on stable nuclei 
has been studied in details since many years, both experimentally and 
theoretically \cite{Primakoff1959,Mukhopadhyay1977,Measday2001,Walecka75}. 
In this process the momentum transfer is of the order of the muon mass and, 
therefore, the calculation of total muon capture rates presents an excellent  
test of models that are also used in studies of low-energy 
neutrino-nucleus reactions. 

In this work we test the fully consistent RHB plus proton-neutron RQRPA model in 
the calculation of total muon capture rates on a large set of nuclei from $^{12}$C to 
$^{244}$Pu, for which experimental values are available \cite{Suzuki1987}. 
Previous calculation of muon capture rates on selected nuclei using the RPA 
approach include the consistent Hartree-Fock (HF) RPA 
model \cite{Auerbach1984,Auerbach1997}, in which the HF mean field and 
the particle-hole interaction result from the same Skyrme effective force, 
and a series of studies \cite{Kolbe1994,Kolbe2000,Zinner2003,Zinner2006} 
in which both the continuum and standard RPA were used, and the effect 
of quenching of axial-vector coupling was analyzed.
The present analysis parallels the recent study by Zinner, Langanke and Vogel 
\cite{Zinner2006}, where the 
nonrelativistic RPA was used to systematically calculate muon capture rates for 
nuclei with $6 \leq Z \leq 94$. There are, however, significant differences between 
the two approaches. The model employed in 
Refs. \cite{Kolbe1994,Kolbe2000,Zinner2003,Zinner2006} uses a 
phenomenological Woods-Saxon potential to generate the basis of 
single-nucleon states. The strength of the potential is adjusted to 
experimental proton and neutron separation energies in individual nuclei.  
In a second step the RPA with a phenomenological Landau-Migdal 
residual interaction is used to calculate 
nuclear excitations. The present approach, as already emphasized above, is fully 
consistent: both the basis of single-nucleon states and multipole excitations of 
nuclei are calculated from the same energy density functional or nuclear 
effective interaction. Results will be compared with data and discussed in relation 
to those reported in Ref.~\cite{Zinner2006}. In particular, we will consider the 
important issue of quenching of the axial-vector strength.

\section{Theoretical framework}

The capture of a negative muon from the atomic $1s$ orbit on a nucleus 
$(Z, N)$
\begin{equation}
\mu^{-} + ({Z},{N}) \longrightarrow  \nu_{\mu} + ({Z-1},{N+1})^{*} \; ,
\end{equation}
presents a simple semi-leptonic reaction that proceeds via the charged 
current of the weak interaction. Detailed expressions for the reaction 
rates and the transition matrix elements can be found in 
Refs.~\cite{O'Connell1972,Walecka75,Walecka2004}. The capture rate reads 
\begin{equation}
\omega_{fi} = \frac{\Omega \nu^{2}}{2\pi} \sum_{\rm lepton\, spins} \frac{1}{2J_{i} + 1}\sum_{M_{i}}\sum_{M_{f}} \left| \langle f\right| \hat{H}_{W} \left| i \rangle \right|^{2}\; ,
\end{equation}
where $\Omega$ denotes the quantization volume
and  $\nu$ is the muon neutrino energy. 
The Hamiltonian $ \hat{H}_{W} $ of the weak interaction 
is expressed in the standard current-current form, i.e. in terms of the 
nucleon $\mathcal{J}_{ \lambda }(\bm{x})$ and lepton $j_{ \lambda }(\bm{x})$ 
currents
\begin{equation}
\hat{H}_{W}=- \frac{G }{ \sqrt{2} }  \int d\bm{x}
\mathcal{J}_{ \lambda }(\bm{x})j^{ \lambda }(\bm{x}) \;,
\end{equation}
and the transition matrix elements read
\begin{equation}
\langle f | \hat{H}_{W}| i \rangle = -  \frac{ G}{\sqrt{2}}l_{ \lambda }
 \int d^3{x}  \frac{\phi_{1s}(\bm{x})}{1/\sqrt{\Omega}}e^{- i \bm{q} \cdot \bm{x}}
\langle f | \mathcal{J}^{ \lambda }(\bm{x}) | i \rangle \; .
\end{equation}
$\phi_{1s}(\bm{x})$ is the muon $1s$ wave function, 
the four-momentum transfer is $q \equiv (q_0, \bm{q})$, and 
the multipole expansion of the leptonic matrix element 
$l_{ \lambda }  e^{- i \bm{q} \cdot \bm{x}}$ determines 
the operator structure for the nuclear transition matrix elements 
\cite{O'Connell1972,Walecka75,Walecka2004}. The expression 
for the muon capture rate is given by 
\begin{equation}
\omega_{fi} = \frac{2 G^{2} \nu^{2}}{(1 + {\nu}/{M_{T}})}~ \frac{1}{2J_{i}+1} 
\left\{ \sum_{J=0}^{\infty} \left| \left\langle J_{f} \left\| \phi_{{1s}} 
\left( \hat{\mathcal{M}}_{J} - \hat{\mathcal{L}}_{J} \right) \right\| J_{i} \right\rangle \right|^{2} + \sum_{J=1}^{\infty} \left| \left\langle J_{f} \left\| \phi_{1s} \left( \hat{\mathcal{T}}^{el}_{J} - \hat{\mathcal{T}}^{mag}_{J} \right) \right\| J_{i} \right\rangle \right|^{2} \right\}
\label{mu-rate}
\end{equation}
where $G$ is the weak coupling constant,  the phase-space factor 
$(1+\nu/M_{T})^{-1}$ accounts for the nuclear recoil, and 
$M_T$ is the mass of the target nucleus. The nuclear transition 
matrix elements between the initial state 
$|J_i \rangle$ and final state $|J_f \rangle$, correspond to the 
charge $\hat{\mathcal{M}}_J$, longitudinal $\hat{\mathcal{L}}_J$, 
transverse electric $ \hat{\mathcal{T}}_J^{EL}$, and 
transverse magnetic $\hat{\mathcal{T}}_J^{MAG}$ multipole operators:
\begin{itemize}
\item the Coulomb operator
\begin{equation}
\hat{\mathcal{M}}_{JM}(\bm{x})= F_1^V M_J^M(\bm{x})
- i \frac{ \kappa  }{m_N} 
\left[  F_A  \Omega _J^M(\bm{x})
+ \frac{ 1}{ 2}  (F_A- m_{\mu} F_P)  \Sigma''^M_J(\bm{x})   \right ]  \;,
\label{coulombop}
\end{equation}
\item the longitudinal operator
\begin{equation}
\hat{\mathcal{L}}_{JM}(\bm{x})=  \frac{q_0}{ \kappa }F_1^V M_J^M(\bm{x})
+i F_A \Sigma ''^M_J(\bm{x}) \;,
\label{longitudinalop}
\end{equation}
\item the transverse electric operator
\begin{equation}
\hat{\mathcal{T}}_{JM}^{el}(\bm{x})=  \frac{\kappa }{ m_N}
 \left[ F^V_1  {\Delta'}_{J}^{M}(\bm{x})+ \frac{1 }{2 } \mu^V
 \Sigma_J^M(\bm{x})     \right ]
+i F_A  {\Sigma'}_J^M (\bm{x}) \;,
\label{transverseelop}
\end{equation}
\item and the transverse magnetic operator
\begin{equation}
\hat{\mathcal{T}}_{JM}^{mag}(\bm{x}) = -i  \frac{ \kappa  }{m_N }
 \left[ F_1^V  \Delta_J^M(\bm{x})- \frac{1 }{2 }  \mu^V  {\Sigma'}_J^M(\bm{x})     \right ]
+F_A  \Sigma_J^M(\bm{x}) \; ,
\label{transversemagop}
\end{equation}
\end{itemize}
where all the form factors are functions of $q^2$, and
$\kappa = \left| \bm{q} \right|$.
These multipole operators contain seven basic
operators expressed in terms of  spherical Bessel functions,
spherical harmonics, and vector spherical harmonics~\cite{O'Connell1972}.
By assuming conserved vector current (CVC), the standard 
set of form factors reads \cite{Kuramoto1990}: 
\begin{equation}
F_{1}^{V} (q^{2}) = \left[ 1 + \left( \frac{q}{840\,\textrm{MeV}} \right)^{2} \right]^{-2},
\label{ff1}
\end{equation}
\begin{equation}
\mu^{V} (q^{2}) = 4.706 \left[ 1 + \left( \frac{q}{840\,\textrm{MeV}} \right)^{2} \right]^{-2},
\label{ff2}
\end{equation}
\begin{equation} \label{eq:fa}
F_{A} (q^{2}) = -1.262 \left[ 1 + \left( \frac{q}{1032\,\textrm{MeV}} \right)^{2} \right]^{-2},
\label{ff3}
\end{equation}
\begin{equation} \label{eq:fp}
F_{P} (q^{2}) = \frac{2 m_{N} F_{A}(q^{2})}{q^{2} + m_{\pi}^{2}} \; .
\label{ff4}
\end{equation}
\bigskip

The muon capture rates are evaluated using Eq.~(\ref{mu-rate}), 
with the transition matrix elements between the initial and final states determined in a
fully microscopic theoretical framework based on the Relativistic Hartree-Bogoliubov (RHB) model for the nuclear ground state, and excited states are calculated using the relativistic quasiparticle random phase approximation (RQRPA).
The RQRPA has been formulated in the canonical single-nucleon basis of the 
relativistic Hartree-Bogoliubov (RHB) model in Ref.~\cite{Paar2003}, and extended
to the description of charge-exchange excitations (proton-neutron RQRPA) 
in Ref.~\cite{Paar2004}. In addition to configurations built from two-quasiparticle
states of positive energy, the relativistic QRPA configuration space must also include
pair-configurations formed from the fully or partially occupied states of
positive energy and empty negative-energy states from the Dirac sea.
The RHB+RQRPA model is fully consistent: in the particle-hole ($ph$) 
channel effective Lagrangians with density-dependent meson-nucleon 
couplings are employed, and pairing ($pp$) correlations are described 
by the pairing part of the finite range Gogny
interaction. Both in the $ph$ and $pp$ channels, the same interactions
are used in the RHB equations that determine the canonical quasiparticle basis, and 
in the matrix equations of the RQRPA. In this work we use one of the 
most accurate meson-exchange 
density-dependent relativistic mean-field effective interactions -- 
DD-ME2~\cite{Lalazissis2005} in the $ph$ channel, and the 
finite range Gogny interaction D1S~\cite{Berger1991} in the $pp$ channel.

The spin-isospin-dependent interaction terms are generated by the $\pi$- and $\rho$-meson exchange. Although the direct one-pion contribution to the nuclear ground state vanishes at the mean-field level because of parity conservation, the pion
must be included in the calculation of spin-isospin excitations. The particle-hole residual interaction of the PN-RQRPA is derived from the
 Lagrangian density:
\begin{equation}
\mathcal{L}^{\textrm{int}}_{\pi + \rho} = -g_{\rho} \bar{\psi}\gamma^{\mu}\vec{\rho}_{\mu}\vec{\tau} \psi - \frac{f_{\pi}}{m_{\pi}} \bar{\psi} \gamma_{5} \gamma^{\mu} \partial_{\mu} \vec{\pi} \vec{\tau} \psi
\end{equation}
where vectors in isospin space are denoted by arrows.
For the density-dependent coupling strength of the $\rho$-meson
to the nucleon we choose the value that is
used in the DD-ME2 effective interaction~\cite{Lalazissis2005}, and
the standard value for the pseudovector pion-nucleon coupling is
${f_{\pi}^{2}}/{4\pi} = 0.08$, and $m_{\pi} = 138$ MeV.
The derivative type of the pion-nucleon coupling
necessitates the inclusion of the zero-range Landau-Migdal term,
which accounts for the contact part of the nucleon-nucleon interaction
\begin{equation}
V_{\delta \pi} = g' \left( \frac{f_{\pi}}{m_{\pi}} \right)^{2} \vec{\tau}_{1} \vec{\tau}_{2} \bsr[\Sigma]_{1} \cdot \bsr[\Sigma]_{2} \delta
(\bsr_{1} - \bsr_{2}),
\end{equation}
with the parameter $g^{\prime}$ adjusted in such a way that the PN-RQRPA
reproduces experimental values of Gamow-Teller resonance (GTR)
excitation energies~\cite{Paar2004}. The precise value depends on
the choice of the nuclear symmetry energy at saturation, and for
the DD-ME2 effective interaction $g^{\prime}$=0.52 has
been adjusted to the position of the GTR in $^{208}$Pb. This value is
kept constant for all nuclides calculated in this work.

In the evaluation of muon capture rates (Eq.~(\ref{mu-rate})), 
for each transition operator $\hat{O}_J$ the matrix elements  
between the ground state of the even-even $(N, Z)$ target nucleus 
and the final state are expressed in terms of single-particle matrix elements 
between quasiparticle canonical states, the corresponding occupation probabilities 
and RQRPA amplitudes:
\begin{equation}
 \langle J_f || \hat{O}_J || J_i \rangle =
 \sum_{pn} 
 \langle p || \hat{O}^J || n \rangle
\left( X_{pn}^{J} u_p v_n - Y_{pn}^{J}v_p u_n \right).
\label{redtrans}
\end{equation}
Transitions between the $|0^+\rangle$ ground state of a spherical even-even
target nucleus and excited states in the corresponding odd-odd nucleus are 
considered. The total muon capture rate is calculated from the expression:
\begin{eqnarray}
& \omega & = 2 G^{2} 
\Bigg\{ \sum_{J_f=0}^{\infty} 
 \frac{\nu_f^{2}}{(1 + {\nu_f}/{M_{T}})}~
 \left| \left\langle J_{f} \left\| \phi_{{1s}} 
\left( \hat{\mathcal{M}}_{J_f} - \hat{\mathcal{L}}_{J_f} \right) \right\| 0^+ \right\rangle \right|^{2}  \nonumber \\
&& + \sum_{J_f=1}^{\infty} 
 \frac{\nu_f^{2}}{(1 + {\nu_f}/{M_{T}})}~
 \left| \left\langle J_{f} \left\| \phi_{1s} \left( \hat{\mathcal{T}}^{el}_{J_f} - \hat{\mathcal{T}}^{mag}_{J_f} \right) \right\| 0^+ \right\rangle \right|^{2} \Bigg\} \; ,
\label{mu-rate-tot}
\end{eqnarray}
with the neutrino energy determined 
by the energy conservation relation
\begin{equation}
m_{\mu} - \epsilon_b +E_i = E_f +\nu_f \; ,
\end{equation}
where $\epsilon_b$ is the binding energy of the muonic atom. 

\begin{figure}
\centerline{
        \includegraphics[scale=0.65,angle=0]{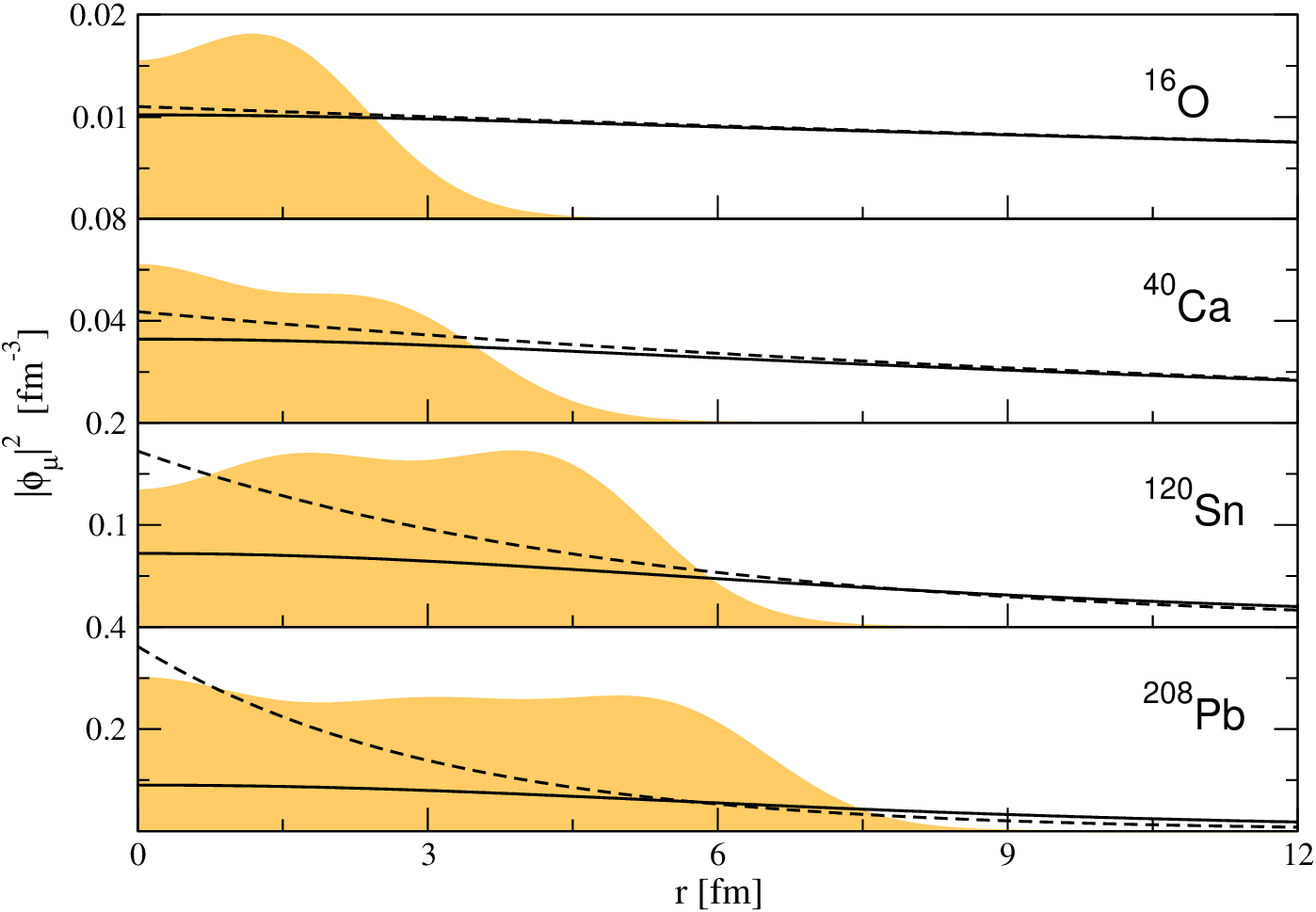}
}
\caption{The square of the $1s$ muon wave function in the 
Coulomb potentials of self-consistent ground-state charge densities 
of $^{16}$O, $^{40}$Ca, $^{120}$Sn and $^{208}$Pb (solid curves), compared 
to eigenfunctions of the Coulomb potential for the corresponding point charge Z 
(dashed curves). The figures also include the calculated charge densities 
of the four nuclei, scaled by arbitrary factors.}
\label{fig:muonwf}
\end{figure}

For each nucleus the muon wave function and binding energy 
are calculated as solutions of 
the Dirac equation with the Coulomb potential determined by the self-consistent 
ground-state charge density. However, while the RHB single-nucleon 
equations are solved by expanding nucleon spinors and meson fields 
in terms of eigenfunctions of a 
spherically symmetric harmonic oscillator potential, the same method could 
not be used for the muon wave functions. The reason, of course, is that the 
muon wave functions extend far beyond the surface of the nucleus and, 
even using a large number of oscillator shells, solutions expressed in terms 
of harmonic oscillator basis functions do not converge.  The Dirac equation 
for the muon is therefore solved in coordinate space using the 
method of finite elements with B-spline shape functions \cite{McNeil1989,deBoor}.  
As an illustration, in Fig. \ref{fig:muonwf} we plot the square of the $1s$ 
muon wave functions for $^{16}$O, $^{40}$Ca, $^{120}$Sn and $^{208}$Pb. 
The solutions that correspond to self-consistent ground-state charge densities 
are compared with eigenfunctions of the Coulomb potential for the corresponding 
point-charge $Z$. For light nuclei the radial dependence of the $1s$ muon wave 
function is not very different from that of the point-charge Coulomb potential.  
With the increase of Z the muon is pulled into the nuclear Coulomb potential, 
and thus the magnitude of the $1s$ density inside the nucleus is reduced 
with respect to the point-charge value. To test our calculation of muon orbitals 
in the nuclear Coulomb potential, in Tables \ref{tab:muon1} and \ref{tab:muon2}
the muon transition energies in Sn isotopes and in $^{208}$Pb, respectively, are 
compared with available data  \cite{Piller1990,Bergem1988}. The calculated 
transition energies are in good agreement with experimental values. 

The effect of the finite distribution of ground-state charge densities on the 
calculated muon capture rates is illustrated in Fig.~\ref{fig:allnuclei}. For 
a large set of nuclei from $^{12}$C to $^{244}$Pu, we plot the ratio between 
calculated and experimental muon capture rates. This ratio is 
$\leq 1.5 $ for all nuclei when the muon $1s$ wave 
functions are determined by self-consistent ground-state charge densities, 
whereas for point-charge Coulomb potentials one notes a distinct increase 
with $Z$, and $\omega_{\rm calc.} / \omega_{\rm exp.} \geq 4$ 
for the heaviest systems.

\section{Results for muon capture rates}

\begin{figure}
\centerline{
        \includegraphics[scale=0.6,angle=0]{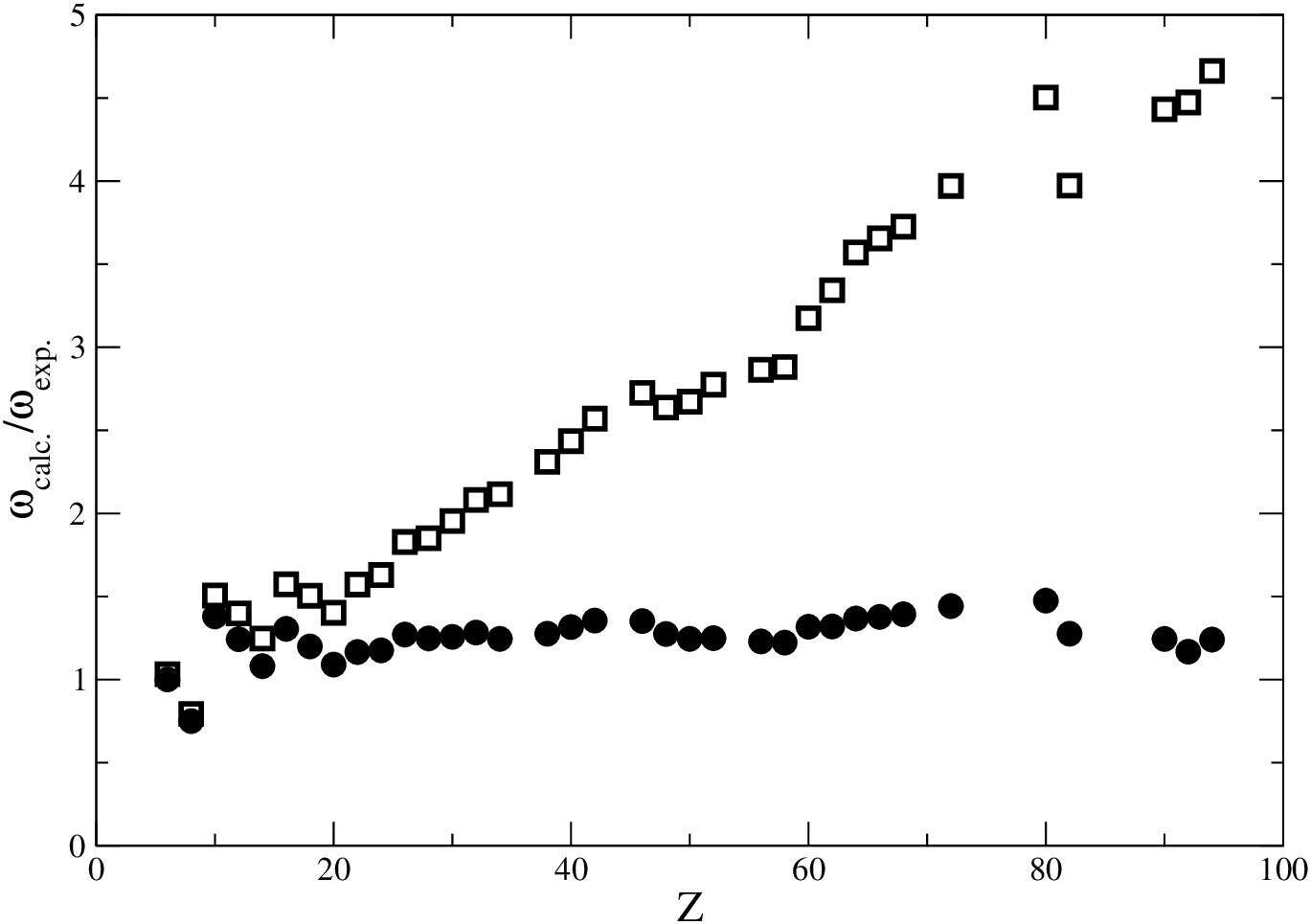}
}
\caption{Ratio of the calculated and experimental total muon capture rates, as 
function of the proton number $Z$. The theoretical values are calculated 
with muon $1s$ wave functions determined by self-consistent ground-state 
charge densities (filled circle symbols), and by the corresponding point-charge 
Coulomb potentials (squares).}
\label{fig:allnuclei}
\end{figure}

The muon capture rates shown in Fig.~\ref{fig:allnuclei} are calculated with
the standard set of free nucleon weak form factors Eqs.~(\ref{ff1}) -- (\ref{ff4}) 
\cite{Kuramoto1990}, i.e. the calculation does not include any in-medium 
quenching of the corresponding strength functions. Even with muon wave 
functions determined self-consistently by  finite-charge densities, the 
resulting capture rates are larger than the corresponding experimental values 
by a factor $\approx 1.2 - 1.4$. This is in contrast to the results of 
Ref.~\cite{Zinner2006}, where the experimental values have been 
reproduced to better than 15\% accuracy, using the free-nucleon weak form 
factors and residual interactions with a mild $A$ dependency. In fact, 
it was shown that the calculated rates for the same residual interactions would be 
significantly below the data if the in-medium quenching of the axial-vector 
coupling constant is employed to other than the true Gamow-Teller (GT) amplitudes. 
Consequently, the calculations reported in Ref.~\cite{Zinner2006} were 
performed with quenching only the GT part of the transition strength 
by a common factor $(0.8)^2 = 0.64$. It was concluded, however, that there 
is actually no need to apply any quenching to operators that contribute 
to the muon capture process, especially those involving single-nucleon 
transitions between major oscillator shells.

As already emphasized in the Introduction, although both calculations are  
based on the RPA framework, there are important differences between the 
model of Ref.~\cite{Zinner2006}, and the RHB+RQRPA approach employed in 
the present study. The main difference is probably the fact that the present 
calculation is fully consistent: for all nuclei both the basis of single-nucleon 
states and the multipole response are calculated using the same effective 
interaction, whereas in Ref.~\cite{Zinner2006} the phenomenological 
Woods-Saxon potential was adjusted to individual nuclei and the strength 
of the residual Landau-Migdal force had a mild $A$-dependence. 

\begin{figure}
\centerline{
        \includegraphics[scale=0.65,angle=0]{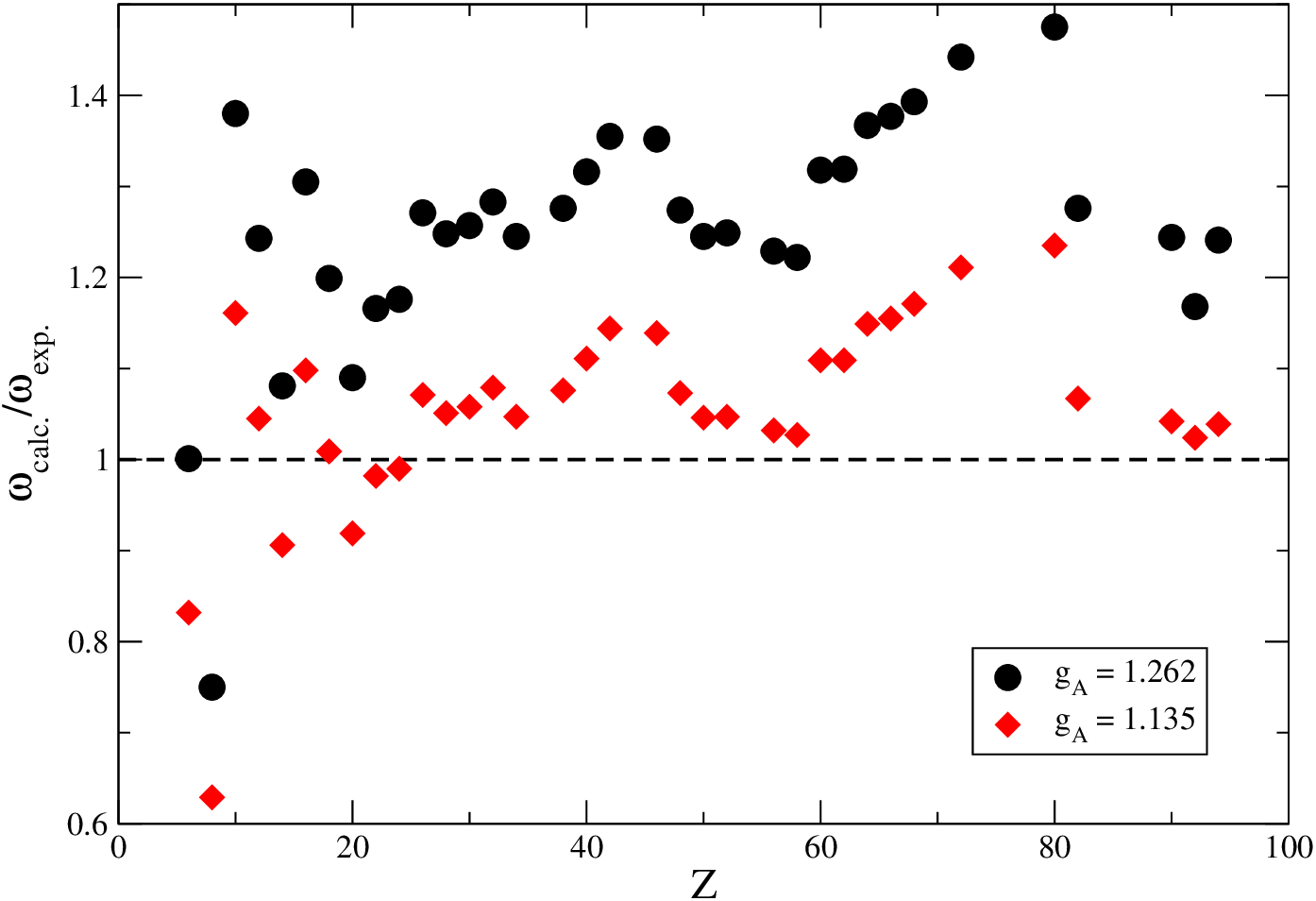}
}
\caption{(Color online) Ratio of the calculated and experimental total muon capture 
rates, as function of the proton number $Z$. Circles correspond to 
rates calculated with the free-nucleon weak form factors 
Eqs.~(\ref{ff1}) -- (\ref{ff4}) \cite{Kuramoto1990}, and diamonds 
denote values obtained by quenching the free-nucleon  axial-vector coupling constant
 $g_A = 1.262$ to $g_A = 1.135$ for all operators, i.e. in all multipole 
channels.}
\label{fig:quenching}
\end{figure}

\begin{figure}
\centerline{
        \includegraphics[scale=0.65,angle=0]{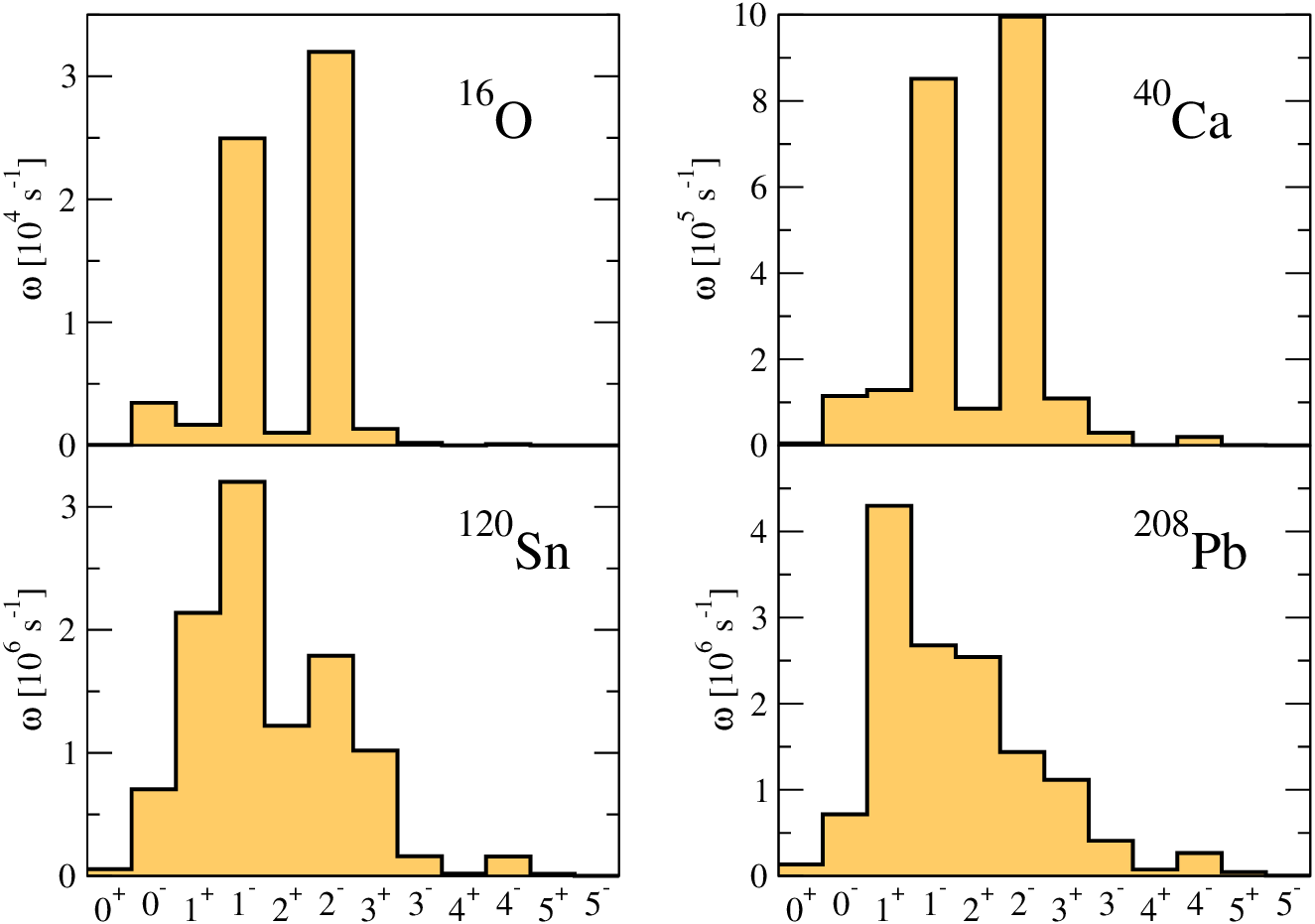}
}
\caption{(Color online) Relative contributions of different multipole transitions to the RHB plus 
RQRPA total muon capture rates in $^{16}$O, $^{40}$Ca, $^{120}$Sn and $^{208}$Pb.}
\label{fig:multipoles}
\end{figure}

In Fig.~\ref{fig:quenching} we compare the ratios of the theoretical 
and experimental total muon capture rates for two sets of weak form factors. 
First, the rates calculated with the free nucleon weak form factors 
Eqs.~(\ref{ff1}) -- (\ref{ff4}) \cite{Kuramoto1990} (circles), and already 
shown in Fig.~\ref{fig:allnuclei}. The lower rates, denoted 
by diamonds, are calculated by applying the same quenching 
$g_A = 1.262 \to g_A = 1.135$ to all axial operators, 
i.e. $g_A$ is reduced by 10\% in all multipole channels. 
In the latter case the level of agreement is very good, with the 
mean deviation between theoretical and experimental values of only 6\%. 
The factor 0.9 with which the free-nucleon  $g_A$ is multiplied is chosen 
in such a way to minimize the deviation from experimental values for 
spherical, closed-shell medium-heavy and heavy nuclei. 
On the average the results are slightly better than those obtained in 
Ref.~\cite{Zinner2006} (cf. Fig. 2 of \cite{Zinner2006}).
Note, however, that in the calculation of Zinner, Langanke and Vogel \cite{Zinner2006} 
only the true Gamow-Teller 
$0\hbar \omega$ transition strength was quenched, rather than the total 
strength in the $1^+$ channel. 
In the present study considerably better results are obtained when the quenched 
value of the axial-vector coupling constants is used for all multipole operators. 
The reason to consider quenching the strength in all multipole channels, rather 
than just for the GT is, of course, that the axial form factor appears in all 
four operators Eqs.~(\ref{coulombop}) -- (\ref{transversemagop}) that induce 
transitions between the initial and final states, irrespective of their multipolarity. 
Even more importantly, only a relatively small contribution to the total capture 
rates actually comes from the GT channel $1^+$. This is illustrated 
in Fig.~\ref{fig:multipoles}, where we display the relative contributions of 
different multipole transitions to the RHB plus RQRPA muon capture rates 
in $^{16}$O, $^{40}$Ca, $^{120}$Sn and $^{208}$Pb. For the two lighter 
$N=Z$ nuclei the dominant multipole transitions are $\lambda^\pi = 1^-$ and 
$\lambda^\pi = 2^-$  (spin-dipole). For the two heavier nuclei there are also 
significant contributions of the $\lambda^\pi = 1^+$ and $\lambda^\pi = 2^+$, 
especially for $^{208}$Pb and for other heavy nuclei. Note that 
in heavy nuclei the $\lambda^\pi = 1^+$ multipole represents $2\hbar \omega$ 
transitions, rather than the $0\hbar \omega$ Gamow-Teller transitions.

Returning to Fig.~\ref{fig:quenching}, we notice that with a 10\% quenching of 
the free-nucleon axial-vector coupling constant $g_A$, 
for medium-heavy and heavy nuclei the calculated capture rates are still 
slightly larger than the corresponding experimental values, with the ratio 
$\omega_{\rm calc.} / \omega_{\rm exp.}$ 
typically around $1.1$, whereas for several lighter  nuclei considered here 
this ratio is actually less than 1 (cf. also Table \ref{tab:allnuclei}). Overall the best 
results, with $\omega_{\rm calc.} / \omega_{\rm exp.} \approx 1$, are obtained 
near closed shells. The characteristic arches between closed shells can probably 
be attributed to deformation effects, not taken into account in our RHB+RQRPA 
model. In addition to the DD-ME2 interaction, we have also carried out a full calculation of 
total capture rates from $^{12}$C to $^{244}$Pu, using the density- and 
momentum-dependent relativistic effective interaction D3C*. In the study 
of $\beta$-decay half-lives of Ref.~\cite{Marketin2006}, this interaction was 
constructed with the aim to enhance the effective (Landau) nucleon mass, 
and thus improve the RQRPA description of $\beta$-decay rates. When D3C* 
is used to calculate muon capture rates, some improvement is obtained only 
locally, for certain regions of $Z$, whereas in other regions ($Z \approx 50$ 
and $Z \geq 82$) the results are not as good as those obtained with DD-ME2. 
The overall quality of the agreement between theoretical and experimental 
capture rates is slightly better with DD-ME2.

\begin{figure}
\centerline{
        \includegraphics[scale=0.65,angle=0]{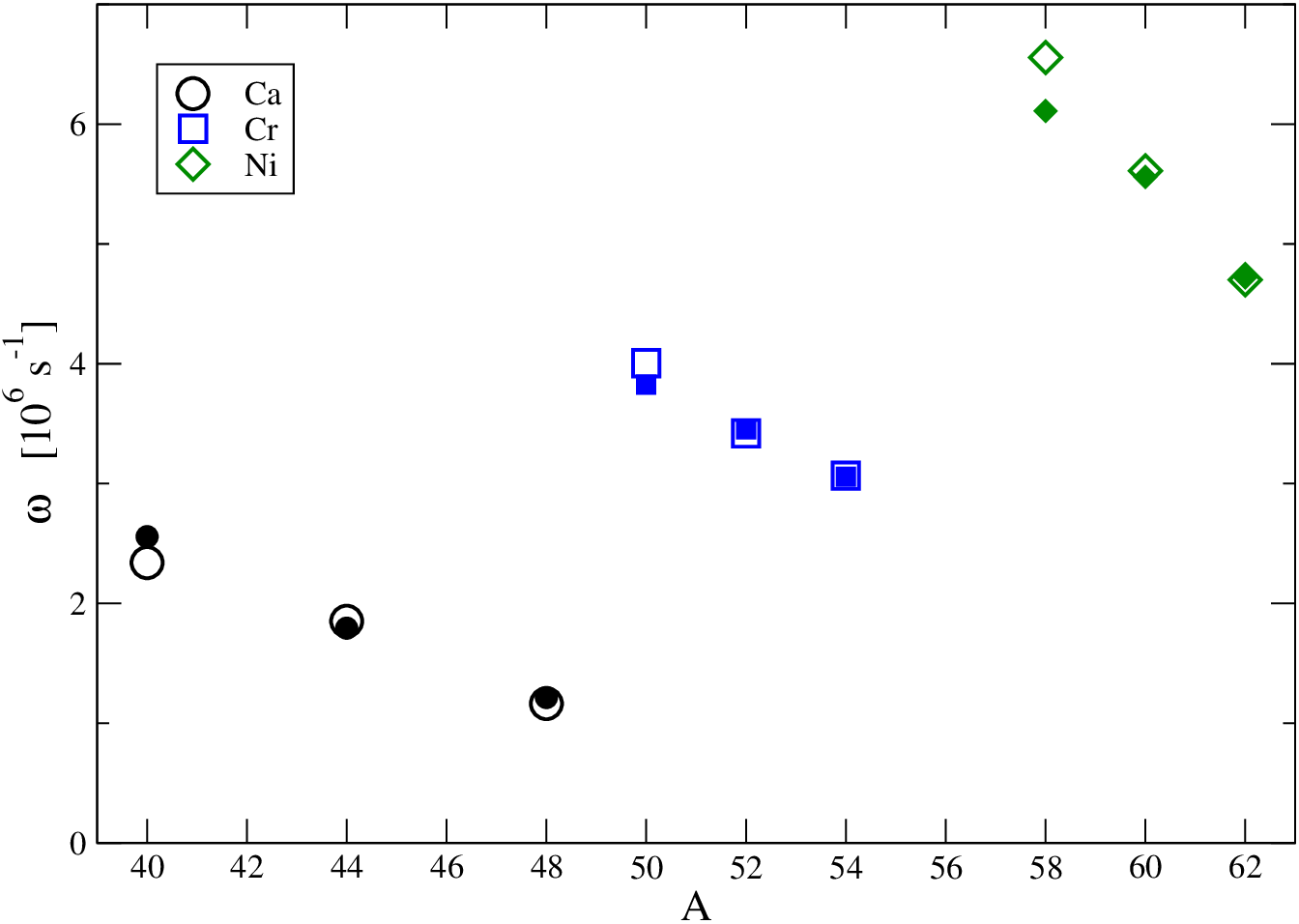}
}
\caption{(Color online) Total muon capture rates on Ca, Cr and Ni isotopes. 
Experimental rates (filled symbols) are compared to theoretical values 
(empty symbols), calculated using the fully 
consistent RHB plus RQRPA framework with the DD-ME2 universal effective interaction, 
and with the quenching of the axial-vector coupling constant 
$g_A = 1.262 \to g_A = 1.135$ for all multipole operators. 
}
\label{fig:cdsn}
\end{figure}

The calculated total muon capture rates for natural elements and individual 
isotopes are also collected in Table \ref{tab:allnuclei}, and compared with 
available data \cite{Suzuki1987}. In particular, the calculation nicely 
reproduces the empirical isotopic dependence of the capture 
rates \cite{Primakoff1959}, i.e. for a given proton number $Z$ the rates 
decrease with increasing neutron number, because of the gradual blocking 
of available neutron levels. The isotopic trend is also illustrated in 
Fig.~\ref{fig:cdsn}, where we plot the experimental and theoretical total 
muon capture rates on Ca, Cr and Ni nuclei. The latter correspond to the 
quenching $g_A = 1.262 \to g_A = 1.135$ for all multipole operators. 

In conclusion, we have tested the RHB plus proton-neutron RQRPA model in 
the calculation of total muon capture rates on a large set of nuclei from $^{12}$C 
to $^{244}$Pu. The calculation is fully consistent, the same universal 
effective interactions are used both in the RHB equations that 
determine the quasiparticle basis, and in the matrix equations of the RQRPA.
The calculated capture rates are sensitive to the in-medium quenching 
of the axial-vector coupling constant. By reducing this constant from its 
free-nucleon value $g_A = 1.262$ to the effective value $g_A = 1.135$ 
for all multipole transitions, i.e. with a quenching of approximately 10\%,
the experimental muon capture rates are reproduced 
with an accuracy better than $10 \%$. This result can be compared to recent 
RPA-based calculations \cite{Kolbe2000,Zinner2003,Zinner2006}, that reproduce 
the experimental values to better than 15\%, using phenomenological potentials 
adjusted to individual nuclei and $A$-dependent residual interactions, but 
without applying any quenching to the operators responsible for the $\mu^-$ 
capture process. The test has demonstrated that the RHB plus 
QRPA model provides a consistent and accurate description of semi-leptonic 
weak interaction processes at finite momentum transfer in medium-heavy and 
heavy nuclei over a large Z-range. The fully consistent microscopic approach, 
based on modern relativistic nuclear energy density functionals, can be extended to 
other types of weak interaction processes (electron capture, neutrino-nucleus 
charge-exchange and neutral-current reactions), and to regions of short-lived 
nuclei far from stability.

\bigskip
\leftline{\bf ACKNOWLEDGMENTS} 
This work was supported by MZOS - project 1191005-1010 and Unity through 
Knowledge Fund (UKF Grant No. 17/08).

\newpage
\clearpage

\clearpage
\clearpage
\clearpage
\clearpage
\clearpage
\begin{table}
\setlength{\tabcolsep}{4.0\tabcolsep}
\caption{\label{tab:muon1} Calculated muon transition energies in tin isotopes
(in units of keV), compared with available data \cite{Piller1990}.}
\begin{tabular}{c|cc|cc} 
\hline
\hline
 & \multicolumn{2}{c|}{$1p_{1/2} - 1s_{1/2}$} & \multicolumn{2}{c}{$1p_{3/2} - 1s_{1/2}$} \\
 & exp. & calc. & exp. & calc. \\
\hline
$^{112}$Sn & 3432 & 3439 & 3478 & 3485 \\
$^{114}$Sn & 3426 & 3432 & 3471 & 3478 \\
$^{116}$Sn & 3420 & 3427 & 3465 & 3472 \\
$^{118}$Sn & & 3421 & & 3466 \\
$^{120}$Sn & 3408 & 3415 & 3454 & 3460 \\
$^{122}$Sn & & 3409 & & 3454 \\
$^{124}$Sn & 3400 & 3404 & 3445 & 3450 \\
\hline
\hline
\end{tabular}
\end{table}

\begin{table}
\setlength{\tabcolsep}{4.0\tabcolsep}
\caption{Calculated muon transition energies in $^{208}$Pb (in units of keV), 
in comparison with experimental values \cite{Bergem1988}.}
\label{tab:muon2}
\begin{center}
\begin{tabular}{c|cc} 
\hline
\hline
$^{208}$Pb  & exp. & calc.\\
\hline
$1p_{3/2} - 1s_{1/2}$  & 5963 & 5956\\
$1p_{1/2} - 1s_{1/2}$  & 5778 & 5773\\
$1d_{3/2} - 1p_{1/2}$  & 2642 & 2633\\
$1d_{5/2} - 1p_{3/2}$  & 2501 & 2493\\
$1d_{3/2} - 1p_{3/2}$  & 2458 & 2450\\
\hline
\hline
\end{tabular}
\end{center}
\end{table}
\begin{table}
\setlength{\tabcolsep}{2.0\tabcolsep}
\caption{Experimental and calculated muon capture rates for natural elements and 
individual isotopes. The theoretical rates are calculated using the fully consistent 
RHB plus RQRPA framework with the DD-ME2 universal effective interaction, 
and with the quenching of the axial-vector coupling constant 
$g_A = 1.262 \to g_A = 1.135$ for all multipole operators. 
Values for naturally occuring elements (element symbol with no superscript) 
are weighted averages of capture rates on individual isotopes, using their natural abundances. Experimental values are from Ref. \cite{Suzuki1987}, unless otherwise stated. 
All rates are in units of $10^{6}\, \textrm{s}^{-1}$.}
\label{tab:allnuclei}
\begin{center}
\begin{tabular}{ccc|ccc|ccc|ccc} 
\hline
\hline
Nucleus & Exp. & Calc. & Nucleus & Exp. & Calc. & Nucleus & Exp. & Calc. & Nucleus & Exp. & Calc. \\
\hline
$^{12}$C  & 0.039 & 0.032 & $^{78}$Se  &       &  6.644 & $^{122}$Sn &       &  9.645 & $^{164}$Dy &       & 13.540 \\
$^{16}$O  & 0.103 & 0.065 & $^{80}$Se  &       &  5.796 & $^{124}$Sn &       &  8.837 & Dy         & 12.29 & 14.194 \\
$^{18}$O  & 0.088 & 0.057 & $^{82}$Se  &       &  4.935 & Sn         & 10.44 & 10.923 & $^{166}$Er &       & 16.129 \\
$^{20}$Ne & 0.204 & 0.237 & Se         & 5.681 &  5.950 & $^{126}$Te &       & 10.652 & $^{168}$Er &       & 14.949 \\
$^{24}$Mg & 0.484 & 0.506 & $^{86}$Sr  &       &  8.885 & $^{128}$Te &       &  9.830 & $^{170}$Er &       & 13.912 \\
$^{28}$Si & 0.871 & 0.789 & $^{88}$Sr  &       &  7.393 & $^{130}$Te &       &  9.068 & Er         & 13.04 & 15.270 \\
$^{32}$S  & 1.352 & 1.485 & Sr         & 7.020 &  7.553 & Te         & 9.270 &  9.706 & $^{178}$Hf &       & 16.434 \\
$^{40}$Ar & 1.355 & 1.368 & $^{90}$Zr  &       &  9.874 & $^{132}$Xe & 9.4$^{b}$ & 10.631 & $^{180}$Hf &       & 15.276 \\
$^{40}$Ca & 2.557 & 2.340 & $^{92}$Zr  &       &  9.694 & $^{136}$Xe & 8.6$^{b}$ &  8.625 & Hf         & 13.03 & 15.783 \\
$^{44}$Ca & 1.793 & 1.851 & $^{94}$Zr  &       &  8.792 & $^{136}$Ba &       & 11.461 & $^{182}$W  &       & 17.259 \\
$^{48}$Ca & 1.214$^{a}$ & 1.163 & Zr   & 8.660 &  9.619 & $^{138}$Ba &       & 10.127 & $^{184}$W  &       & 15.938 \\
$^{48}$Ti & 2.590 & 2.544 & $^{92}$Mo  &       & 12.374 & Ba         & 9.940 & 10.259 & $^{186}$W  &       & 14.807 \\
$^{50}$Cr & 3.825 & 4.001 & $^{94}$Mo  &       & 12.001 & $^{140}$Ce &       & 11.888 & W          & 12.36 & 15.971 \\
$^{52}$Cr & 3.452 & 3.419 & $^{96}$Mo  &       & 10.933 & $^{142}$Ce &       & 12.142 & $^{198}$Hg &       & 17.369 \\
$^{54}$Cr & 3.057 & 3.065 & $^{98}$Mo  &       &  9.804 & Ce         & 11.60 & 11.917 & $^{200}$Hg &       & 16.227 \\
Cr        & 3.472 & 3.483 & Mo         & 9.614 & 10.995 & $^{142}$Nd &       & 14.043 & $^{202}$Hg &       & 15.205 \\
$^{56}$Fe & 4.411 & 4.723 & $^{104}$Pd &       & 13.182 & $^{144}$Nd &       & 14.288 & $^{204}$Hg &       & 13.993 \\
$^{58}$Ni & 6.110 & 6.556 & $^{106}$Pd &       & 11.912 & $^{146}$Nd &       & 12.981 & Hg         & 12.74 & 15.733 \\
$^{60}$Ni & 5.560 & 5.610 & $^{108}$Pd &       & 10.795 & Nd         & 12.50 & 13.861 & $^{206}$Pb &       & 15.717 \\
$^{62}$Ni & 4.720 & 4.701 & $^{110}$Pd &       &  9.821 & $^{148}$Sm &       & 15.425 & $^{208}$Pb &       & 13.718 \\
Ni        & 5.932 & 6.234 & Pd         & 10.00 & 11.391 & $^{150}$Sm &       & 14.132 & Pb         & 13.45 & 14.348 \\
$^{64}$Zn &       & 6.862 & $^{110}$Cd &       & 12.960 & $^{152}$Sm &       & 13.451 & $^{232}$Th & 12.56 & 13.092 \\
$^{66}$Zn &       & 5.809 & $^{112}$Cd &       & 11.800 & $^{154}$Sm &       & 12.563 & $^{234}$U  & 13.79 & 14.231 \\
$^{68}$Zn &       & 4.935 & $^{114}$Cd &       & 10.746 & Sm         & 12.22 & 13.554 & $^{236}$U  & 13.09$^{c}$ & 13.490 \\
Zn        & 5.834 & 6.174 & $^{116}$Cd &       &  9.829 & $^{156}$Gd &       & 14.785 & $^{238}$U  & 12.57$^{c}$ & 12.872 \\
$^{70}$Ge &       & 6.923 & Cd         & 10.61 & 11.381 & $^{158}$Gd &       & 13.573 & $^{242}$Pu & 12.90 & 13.554 \\
$^{72}$Ge &       & 5.970 & $^{116}$Sn &       & 12.395 & $^{160}$Gd &       & 12.460 & $^{244}$Pu & 12.40$^{d}$ & 12.887 \\ 
$^{74}$Ge &       & 5.519 & $^{118}$Sn &       & 11.369 & Gd         & 11.82 & 13.580 & & & \\
Ge        & 5.569 & 6.011 & $^{120}$Sn &       & 10.486 & $^{162}$Dy &       & 14.917 & & & \\
\hline
\hline
\end{tabular}
\end{center}
\begin{flushleft}
$^{a}$ From Ref. \cite{Fynbo2003}. \\
$^{b}$ From Ref. \cite{Mamedov2000}. \\
$^{c}$ From Ref. \cite{Haenscheid1990}. \\
$^{d}$ From Ref. \cite{David1988}. \\
\end{flushleft}
\end{table}
\end{document}